# Structural variations and magnetic properties of the quantum antiferromagnets $Cs_2CuCl_{4-x}Br_x$

Pham Thanh Cong[1], Bernd Wolf[1], Natalija van Well[1], Amir A. Haghighirad[1,2], Franz Ritter[1]
Wolf Assmus[1], Cornelius Krellner[1] and Michael Lang[1]

[1]Physics Institute, Goethe-University Frankfurt (M), SFB/TR 49, D-60438 Frankfurt (M), Germany
[2]Clarendon Laboratory, University of Oxford, Parks Road, Oxford OX1 3PU, United Kingdom

**Depending on the crystal growth conditions, an orthorhombic (O-type) or a tetragonal (T-type) structure can be found in the solid solution $Cs_2CuCl_{4-x}Br_x$ ($0 \leq x \leq 4$). Here we present measurements of the temperature-dependent magnetic susceptibility and isothermal magnetization on the T-type compounds x = 1.6 and 1.8 and compare these results with the magnetic properties recently derived for the O-type variant by Cong et al., Phys. Rev. B 83, 064425 (2011). The systems were found to exhibit quite dissimilar magnetic properties which can be assigned to differences in the Cu coordination in these two structural variants. Whereas the tetragonal compounds can be classified as quasi-2D ferromagnets, characterized by ferromagnetic layers with a weak antiferromagnetic inter-layer coupling, the orthorhombic materials, notably the border compounds x = 0 and 4, are model systems for frustrated 2D Heisenberg antiferromagnets.**

## I. INTRODUCTION

Low-dimensional (low-D) quantum magnets reveal a wealth of fascinating and unexpected phenomena. Of particular interest are the anomalous properties which result from the interplay of strong quantum fluctuations and geometric frustration [1] – [3] which may give rise to spin-liquid behavior at low temperatures. The simplest model for such a scenario is the spin $S = ½$, 2D triangular-lattice Heisenberg antiferromagnet [4] – [6]. A good realization of the spatially anisotropic version of this model is provided by the orthorhombic layered compounds $Cs_2CuCl_4$ [7], [8] and $Cs_2CuBr_4$ [9]. For $Cs_2CuCl_4$, the frustration effects are derived from a dominant antiferromagnetic exchange coupling $J/k_B$ = 4.34(6) K [10] along the in-plane $b$-axis together with a second in-plane coupling $J' \sim J/3$ along a diagonal bond in the $bc$-plane [11]. Other magnetic couplings in this material, like the anisotropic Dzyaloshinskii-Moriya interaction and the inter-plane interaction $J''$, are more than an order of magnitude smaller than $J$ [11]. The finite inter-layer coupling eventually leads to long-range antiferromagnetic order at $T_N$ = 0.62 K and a field-induced quantum-critical point (QCP) at $B_c$ ~ 8.5 T ($B//a$) for $Cs_2CuCl_4$. For the isostructural compound, $Cs_2CuBr_4$, the coupling constant $J$ and the degree of frustration, $J'/J = 0.74$, are larger in comparison to the chlorine system. The inter-layer coupling $J''$ in $Cs_2CuBr_4$ leads to a Néel ordering at $T_N$ = 1.42 K with a corresponding QCP at $B_c$ ~ 32 T ($B//a$). These border cases thus motivated the study of the magnetic properties of the $Cs_2CuCl_{4-x}Br_x$ solid solution [12], in which, by a continuous replacement of $Cl^-$ by $Br^-$, frustration effects were expected to become increasingly important. Much to our surprise, however, we have found a discontinuous evolution of the magnetic properties of $Cs_2CuCl_{4-x}Br_x$ with x. Our study revealed three distinct magnetic regimes, with border-line concentrations $x_{c1}$ = 1 and $x_{c2}$ = 2 marking particularly interesting cases for the O-type structure. In addition, depending on the condition of the crystal growth, a tetragonal structure [13] was found. In this contribution, we present the magnetic characterization of this new T-type material which, together with our previous results on the O-type structure [12], provides the basis for a better understanding of the structure-magnetic properties relations in the Cu-halide systems.

## II. CRYSTAL GROWTH

High-quality single crystals of the $Cs_2CuCl_{4-x}Br_x$ ($0 \leq x \leq 4$) mixed system were grown from aqueous solutions by an evaporation technique; see [13] for details.

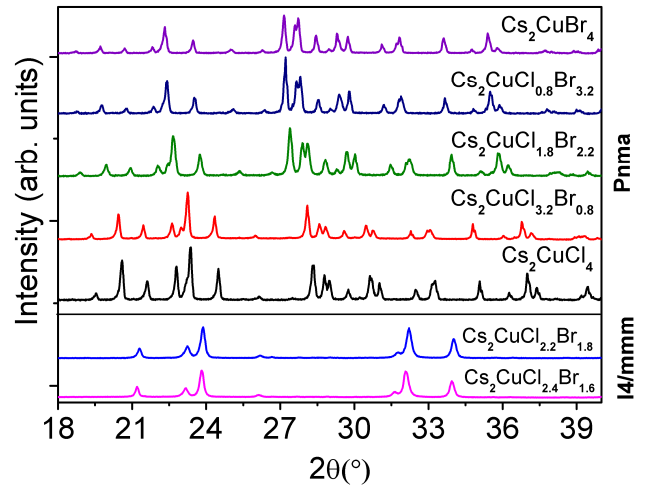

Fig. 1. Results of powder X-ray diffraction (XRD) measurements on $Cs_2CuCl_{4-x}Br_x$ with selected compositions (Cu Kα radiation) of the orthorhombic (x = 0, 0.8, 2.2, 3.2, 4) and the tetragonal (x = 1.6, 1.8) structure.

This method is preferred over simple temperature reduction for initiating and maintaining the growth process, because, depending on the temperature during the crystal growth, either the O-type orthorhombic (*Pnma*) or the T-type tetragonal (*I4/mmm*) structure can be realized for a given composition. For growth temperatures of at least 50 °C, the orthorhombic structure of the end members of the mixed crystal series is preserved over the whole composition range. This can be seen in fig. 1 where selected powder diffractograms are shown for $Cs_2CuCl_{4-x}Br_x$ with different Br contents. Vegard's law gives a first approximation for the variation of the unit cell volume with Br content for the O-type material, cf. the black line in fig. 2.

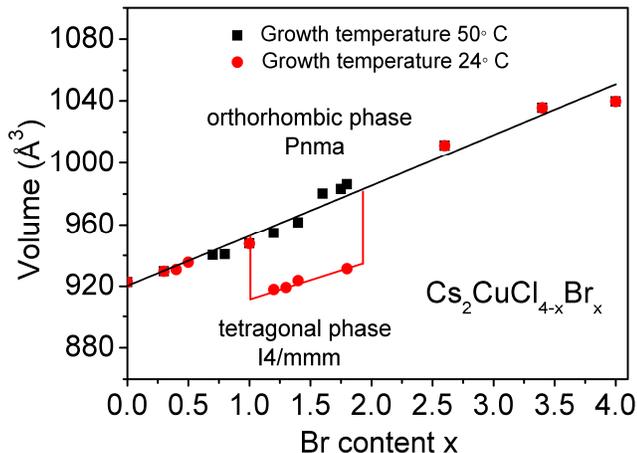

Fig. 2. Unit cell volume vs Br concentration x. For growth at 50 °C (black line), the volume increases linearly with increasing Br content whereas for growth at 24 °C (red line), the volume is significantly reduced within the existence range of the tetragonal modification.

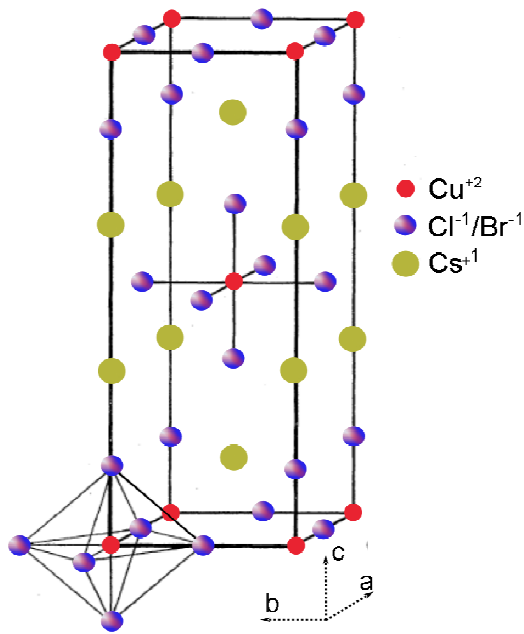

Fig. 3. Room temperature crystal structure of tetragonal $Cs_2CuCl_{4-x}Br_x$. The $Cu^{2+}$ ions are surrounded by tetragonally deformed halide octahedra elongated in the *ab* plane.

If the growth of $Cs_2CuCl_{4-x}Br_x$ takes place at room temperature, within ~1 ≤ x ≤ ~2, no crystals belonging to the orthorhombic structure of $Cs_2CuCl_4$ and $Cs_2CuBr_4$ are formed. Instead, a tetragonal phase is obtained as verified by powder XRD (X-ray diffraction) measurements, cf. fig. 1 for the spectra of the x = 1.6, 1.8 samples and fig. 2 (red line) for the variation of the unit-cell volume with varying Br content. The exact borders of the concentration range for the T-type variant seem to be slightly dependent on the growth temperature [13]. In this phase, the $Cu^{2+}$ ions are surrounded by deformed halide octahedra which are elongated in the *ab* plane. The deformation of the octahedra is assigned to the Jahn-Teller effect. At room temperature the T-type phase can be described by the $K_2NiF_4$ structure type with space group *I4/mmm* as shown in fig. 3. Note that the T-type structure of $Cs_2CuCl_{4-x}Br_x$ (1 ≤ x ≤ 2) is not related to the tetragonal phase of $Cs_2CuCl_4 \cdot 2H_2O$ (space group $P4_2/mnm$), well-known in literature [14].

### III. MAGNETIC CHARACTERIZATION

The magnetic susceptibility was measured in the temperature range 2 K ≤ T ≤ 300 K and in various magnetic fields B ≤ 1 T using a Quantum Design SQUID magnetometer. The magnetization at constant temperature was determined up to a field of 5 T. The data were corrected for the temperature-independent diamagnetic core contribution, according to [15], and the magnetic contribution of the sample holder. The latter was determined from an independent measurement. A careful determination of the magnetic background is important, especially at high temperatures where the materials' magnetic signal is small.

### IV. RESULTS AND DISCUSSION

Figure 4a shows a compilation of the results for the molar magnetic susceptibility $\chi_{mol}(T)$ for the $Cs_2CuCl_{4-x}Br_x$ (0 < x < 4) mixed system at temperatures 2 K ≤ T ≤ 40 K including the pure border compounds (x = 0 and x = 4.0). The $\chi_{mol}(T)$-values of the T-type tetragonal compounds with x = 1.6 and x = 1.8 are shown on the left scale whereas the values for the O-type systems are displayed on the right scale. In the temperature range 30 K ≤ T ≤ 300 K, the $\chi_{mol}(T)$ data for all materials investigated follow a Curie-Weiss-type temperature dependence.

The susceptibility curves of the pure compounds (x = 0 and 4.0) in the O-type phase reveal a continuous increase with decreasing temperature and a pronounced maximum at $T_{max}$ = (2.8 ± 0.15) K for $Cs_2CuCl_4$ and (8.8 ± 0.1) K for $Cs_2CuBr_4$, which is distinctly broader for the latter compound due to the larger J'/J ratio of 0.74 [16] as compared to J'/J = 0.34 for $Cs_2CuCl_4$ [11]. The maximum reflects the low-dimensional magnetic character of both materials in the temperature range under investigation. As shown in [12], a parameterization of the susceptibility data for various concentrations with respect to the position of the susceptibility maximum and its height, $T_{max}$ and $\chi_{mol}(T_{max})$, respectively, shows that there is no continuous evolution of the magnetic properties as a function of the Br concentration. Instead, three distinct concentration regimes could be identified, which are separated by critical concentrations $x_{c1} = 1$ and $x_{c2} = 2$. This unusual magnetic behavior of the O-type structure could be explained by considering the structural peculiarities of the materials, which support a site-selective replacement of Cl$^-$ by Br$^-$ ions [12]. All three concentration regimes have in common that the magnetic properties are those of quasi-2D Heisenberg antiferromagnets. In contrast, the T-type compounds, with a Br content lying in-between the critical concentrations $x_{c1} = 1$ and $x_{c2} = 2$, exhibit a huge increase in the susceptibility below 30 K, cf. fig. 4a. In the data for the concentration x = 1.8, taken at a field of 1 T, we find a kink-like anomaly at $T_N = 8.5$ K which we attribute to a phase transition into long-range antiferromagnetic order. For x = 1.6 at the same field of 1 T we get $T_N = 5.6$ K. The susceptibility is nearly temperature independent within the antiferromagnetically ordered phase for the temperature range investigated.

compounds exhibit a value for $n_{eff}$ at 300 K which is very close to the value of the free Cu$^{2+}$ ion. The important result of fig. 4b is the totally different temperature dependence of $n_{eff}$ for the O- and T-type structures. For both compounds in the orthorhombic structure, $n_{eff}$ is decreasing continuously with decreasing temperature. Such a behavior is expected for materials where the Cu$^{2+}$ ions are in a tetrahedral environment. The strong reduction of $n_{eff}$ at low temperatures (below 40 K) is due to the dominant antiferromagnetic exchange in both materials. In contrast, for the T-type structure, $n_{eff}$ is increasing with decreasing temperature down to the temperature of the antiferromagnetic order, reflecting a dominant ferromagnetic exchange for temperatures above $T_N$. As a consequence of long-range antiferromagnetic order among the ferromagnetic planes, the effective magnetic moment becomes strongly suppressed for $T < T_N$.

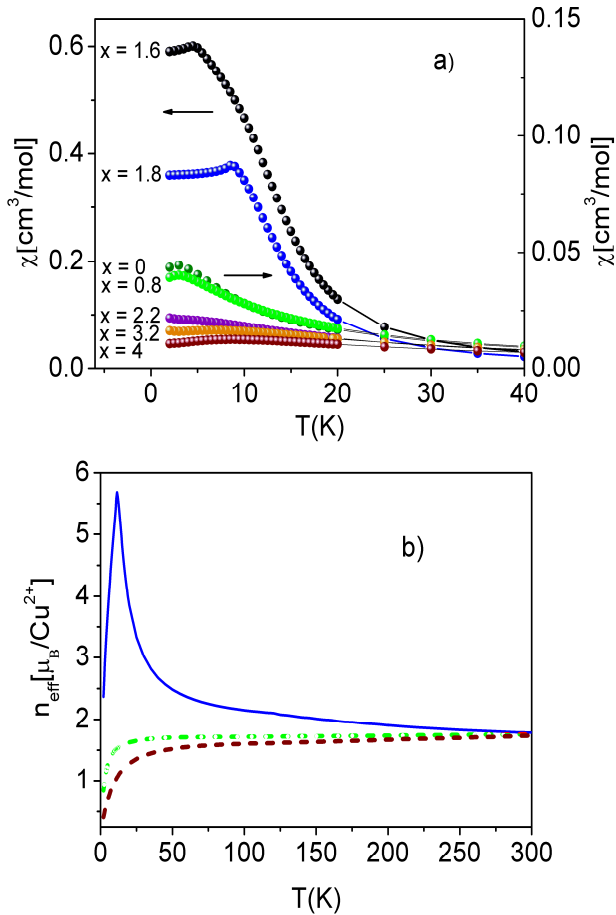

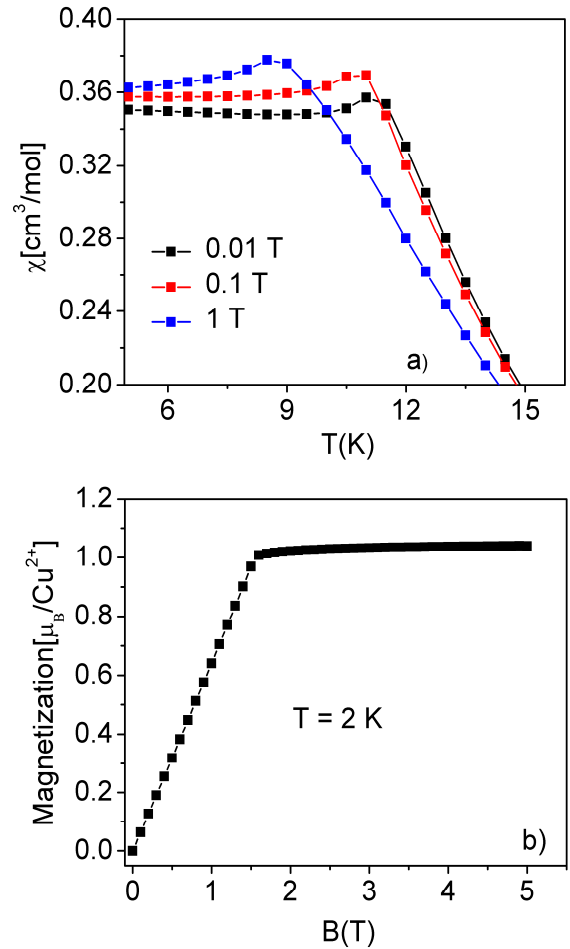

Fig. 4. a) Molar susceptibility of single crystalline Cs$_2$CuCl$_{4-x}$Br$_x$ ($0 \leq x \leq 4$) for the T-type phase (x = 1.6 and 1.8) (left scale) and the O-type phase (x = 0, 0.8, 2.2, 3.2, 4.0 taken from [12]) (right scale) as a function of temperature. b) Effective magnetic moment of O-type Cs$_2$CuCl$_4$ (green broken line) and O-type Cs$_2$CuBr$_4$ (brown dotted line) together with the results for the T-type Cs$_2$CuCl$_{2.2}$Br$_{1.8}$ (blue solid line).

Fig. 5. a) $\chi_{mol}(T)$ for the T-type phase of Cs$_2$CuCl$_{2.2}$Br$_{1.8}$ for magnetic fields of 0.01 T (black squares), 0.1 T (red squares) and 1.0 T (blue squares). b) Isothermal magnetization at 2 K for fields up to 5 T for $B // c$.

The fig. 4b shows the effective magnetic moment $n_{eff}$, defined as $n_{eff} = g [S (S + 1)]^{1/2}$, of O-type Cs$_2$CuCl$_4$ (green dashed line) and O-type Cs$_2$CuBr$_4$ (brown dashed line) together with T-type Cs$_2$CuCl$_{2.2}$Br$_{1.8}$ (blue solid line). All three

The magnetic behavior for the T-type Cs$_2$CuCl$_{4-x}$Br$_x$ ($1 < x < 2$) mixed system can be understood by considering the structural properties of the materials. In the T-type compounds the Cu ions are in an octahedral environment and at room temperature the crystal structure is of the K$_2$NiF$_4$ structure type with space group I4/mmm. In analogy to K$_2$CuF$_4$ we anticipate that the octahedra in T-type $Cs_2CuCl_{4-x}Br_x$ are Jahn-Teller distorted and elongated within the tetragonal plane. This would be possible when the halide atoms, located in between the Cu ions along the *a*- and *b*-axis (see fig. 3), are shifted inside the unit cell as displayed in fig. 1 of [17] for $K_2CuF_4$. Such a transition, corresponding to a cooperative orbital ordering, would directly influence the magnetic exchange. We speculate that below the transition all Cu-halide octahedra are oriented such that their elongated axes lie within the tetragonal plane and are perpendicular to each other for neighbouring octahedra. This implies that T-type $Cs_2CuCl_{4-x}Br_x$ ($1 < x < 2$) features $Cu^{2+}$ ions with alternating $d_{z^2-x^2}$ and $d_{z^2-y^2}$ orbitals which, consequently, provide a dominant ferromagnetic exchange within the *ab* planes. An identical scenario is indicated in fig. 1 of [17] for $K_2CuF_4$. Materials, where the magnetic coupling between those tetragonal planes is weak, are described as quasi-2D ferromagnets. Prominent examples for such a scenario include the model systems $K_2CuF_4$ or $Rb_2CuCl_4$ [18].

To obtain additional information about the long-range ordered antiferromagnetic phase, measurements of the magnetic susceptibility at varying magnetic fields were performed. Fig. 5a exhibits $\chi_{mol}(T)$ data for magnetic fields of 0.01 T, 0.1 T and 1.0 T. From these measurements the initial slope of the phase boundary between the high-*T* paramagnetic and low-*T* antiferromagnetic phases can be constructed. The transition temperatures, determined from the position of the kink in the susceptibility, are $T_N = (11.40 \pm 0.01)$ K at $B = 0.01$ T and $T_N = (11.14 \pm 0.01)$ K at 0.1 T, reflecting an almost vertical slope of the phase boundary at this field level. At $B = 1$ T, however, $T_N$ is already significantly reduced to $(8.44 \pm 0.02)$ K which indicates a rather small critical field at low temperatures. Fig. 4b exhibits results of the isothermal magnetization $M$ at $T = 2$ K. $M (B, T = 2$ K$)$ increases linearly with field until the saturation field is reached at $B_s = 1.47$ T. The saturation magnetization corresponds to a magnetic moment of 1.09 $\mu_B$.

For a typical antiferromagnet with a $T_N$ of order 10 K, a critical field of only 1.5 T is rather small. However, such a behavior is usually found in the class of quasi-2D ferromagnets characterized by a weak antiferromagnetic inter-layer coupling, resulting in an antiferromagnetic order of ferromagnetic layers as displayed in Fig. 1b of [18].

## V. Conclusion and outlook

In summary, we have determined the magnetic properties of the solid solution $Cs_2CuCl_{4-x}Br_x$ ($0 \leq x \leq 4$) in the tetragonal phase and compared them to previous results [12] of the orthorhombic variant. For the two structural variants the Cu environment is different, leading to totally different magnetic behavior. The orthorhombic materials, especially the pure Cl- ($x = 0$) and Br- ($x = 4$) compounds, are model systems for frustrated 2D Heisenberg antiferromagnets whereas the tetragonal compounds belong to the class of quasi-2D ferromagnets with a very small antiferromagnetic inter-layer coupling. We speculate that the ferromagnetic intra-layer exchange is due to an orbital ordering, in analogy to the scenario suggested [17] and experimentally verified [19] for $K_2CuF_4$. Still an open question for T-type $Cs_2CuCl_{4-x}Br_x$ ($1 < x < 2$) is the temperature at which the cooperative Jahn-Teller effect (orbital ordering) takes place. This will be the subject of forthcoming structural and thermodynamic investigations.